# Injection and extraction magnets: kicker magnets

*M.J. Barnes*, L. Ducimetière, T. Fowler, V. Senaj, L. Sermeus
CERN, Geneva, Switzerland

**Abstract**
Each stage of an accelerator system has a limited dynamic range and therefore a chain of stages is required to reach high energy. A combination of septa and kicker magnets is frequently used to inject and extract beam from each stage. The kicker magnets typically produce rectangular field pulses with fast rise- and/or fall-times, however, the field strength is relatively low. To compensate for their relatively low field strength, the kicker magnets are generally combined with electromagnetic septa. The septa provide relatively strong field strength but are either DC or slow pulsed. This paper discusses injection and extraction systems with particular emphasis on the hardware required for the kicker magnet.

## 1   Introduction

An accelerator stage has limited dynamic range: a chain of accelerator stages is required to reach high energy. Thus beam transfer into (injection) and out of (extraction) an accelerator is required. The design of the injection and extraction systems aims to achieve the following:

– minimize beam loss,

– place the newly injected or extracted particles onto the correct trajectory, with the correct phase space parameters.

A combination of septa and kickers is frequently used for injection and extraction. Septa can be electrostatic or magnetic: they provide slower field rise- and fall-times, but stronger field, than kicker magnets. Some septa are designed to be operated with DC. Kicker magnets provide fast field rise- and fall-times, but relatively weak fields.

In general, a septum (plural: septa) is a partition that separates two cavities or spaces. In a particle accelerator a septum is a device which separates two field regions. Important features of septa are an ideally homogeneous (electric or magnetic) field in one region, for deflecting beam, and a low fringe field (ideally zero magnetic and electric field) next to the septum so as not to affect the circulating beam. Hence a septum provides a space separation of circulating and injected/extracted beam. In contrast a kicker magnet provides time selection [separation] of beam to be injected/extracted: a kicker system is used for fast, single-turn, injection and extraction.

The processes of injection and extraction are covered in the proceedings of this CERN Accelerator School, in the paper *Injection and extraction magnets: septa*; the aforementioned paper also discusses the hardware associated with septa. The present paper discusses the hardware associated with kicker magnets.

The field produced by a kicker magnet must rise/fall within the time period between the beam bunches (see Sections 2 and 3). In addition, the magnetic field must not significantly deviate from the flat top of the pulse or from zero between pulses (i.e., very small ripple/excursions). Typical field rise/fall-times range from tens to hundreds of nanoseconds and pulse widths range from tens of nanoseconds to tens of microseconds. If a kicker exhibits a time-varying structure, in the field pulse shape, this can translate into small offsets with respect to the closed orbit (betatron oscillations). Thus a fast, low-ripple, kicker system is generally required.

## 2 Single-turn (fast) injection

Figure 1 shows an example of fast single-turn injection in one plane. The injected beam passes through the homogeneous field region (gap) of the septum: circulating beam is in the field-free region (i.e., space separation of injected and circulating beam). The septum deflects the injected beam onto the closed orbit at the centre of the kicker magnet; the kicker magnet compensates the remaining angle. The septum and kicker are either side of a quadrupole (defocusing in the injection plane) which provides some of the required deflection and minimizes the required strength of the kicker magnet.

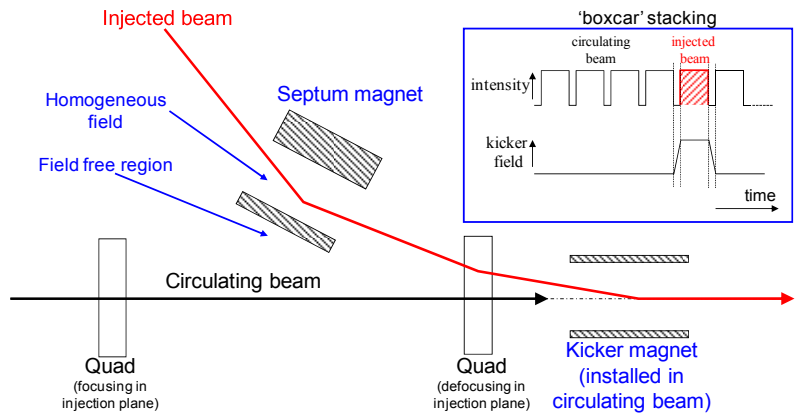

**Fig. 1:** Fast single-turn injection in one plane

A kicker magnet is installed in the accelerator and hence the circulating beam is in the aperture of the kicker. Thus the kicker field must rise from zero to full field in the time interval between the circulating beam and the start of the injected beam (Fig. 1, top right) and fall from full field to zero field in the time interval between the end of the injected beam and the subsequent circulating beam (Fig. 1, top right). The kicker system is described in more detail in Section 4.

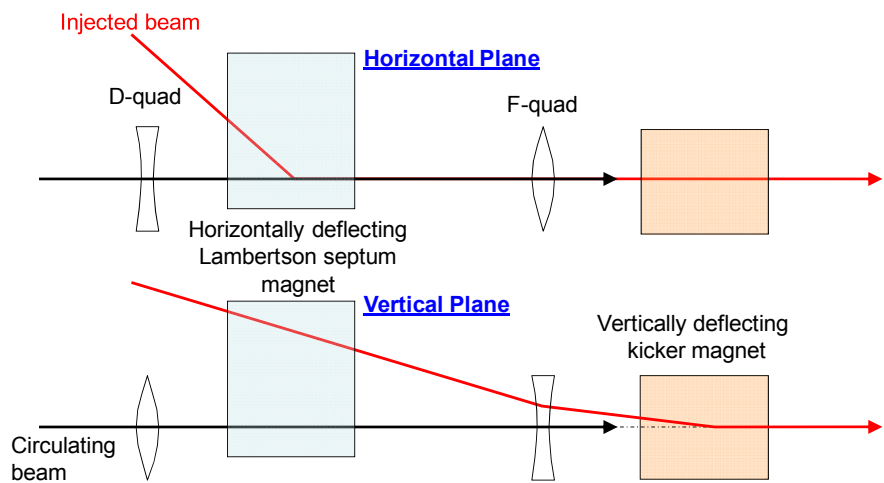

**Fig. 2:** Fast single-turn injection in two planes

Figure 2 shows an example of fast single-turn injection in two planes: a Lambertson septum is used for a two-plane injection scheme. The injected beam passes through the homogeneous field region of the septum: circulating beam is in the field-free region of the septum. In the example shown in Fig. 2 the septum deflects the beam horizontally and the downstream kicker magnet deflects the beam vertically onto the closed orbit of the circulating beam. The septum and kicker are either side of an F-quadrupole (horizontally focusing but vertically defocusing) to minimize the required strength of the kicker magnet. The Lambertson septum magnet is discussed in more detail in the proceedings of this CAS, in Section 5.2.4 of *Injection and extraction magnets: septa*.

## 3  Single-turn (fast) extraction

Extraction is the process of ejecting a particle beam from an accelerator and into a transfer line or a beam dump, at the appropriate time, while minimizing beam loss and placing the extracted particles onto the correct trajectory, with the correct phase space parameters. Extraction usually occurs at higher energy than injection, hence stronger elements (larger $\int B.dl$) are required. At high energies many kicker and septum modules may be needed. To reduce the required strength of the kicker magnet, a closed orbit bump can be applied to bring the circulating beam near to the septum.

Figure 3 shows an example of fast single-turn extraction in one plane. The kicker magnet deflects the entire beam into the septum in a single turn [time selection (separation) of beam to be extracted]. The extracted beam passes through the homogeneous field region of the septum: the circulating beam, prior to extraction, is in the field-free region of the septum (space separation of circulating and extracted beam). The septum deflects the entire kicked beam into the transfer line.

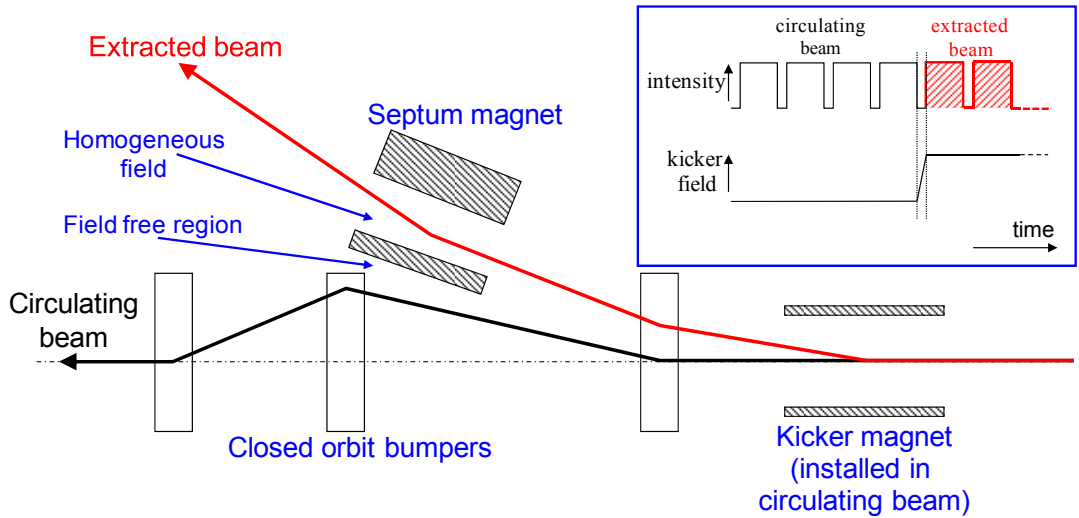

**Fig. 3:** Fast single-turn extraction in one plane

The kicker magnet is installed in the accelerator and hence the circulating beam is in the aperture of the kicker. Thus the kicker field must rise from zero to full field in a beam-free time interval deliberately created in the circulating beam (Fig. 3, top right). The entire beam is generally extracted and hence fast fall-time is typically not required: however, sometimes, bunch-by-bunch transfers are made and then the field of the kicker magnets must have fast rise- and fall-times [1].

## 4 Kicker system

### 4.1 Overview

Figure 4 shows a simplified schematic of a kicker system. The main sub-systems ('components') of a kicker system are

- PFL = Pulse Forming Line (coaxial cable) or PFN = Pulse Forming Network (lumped elements),
- kicker magnet,
- fast, high power, switch(es),
- RCPS = Resonant Charging Power Supply,
- transmission line(s) [coaxial cable(s)],
- terminators (resistive).

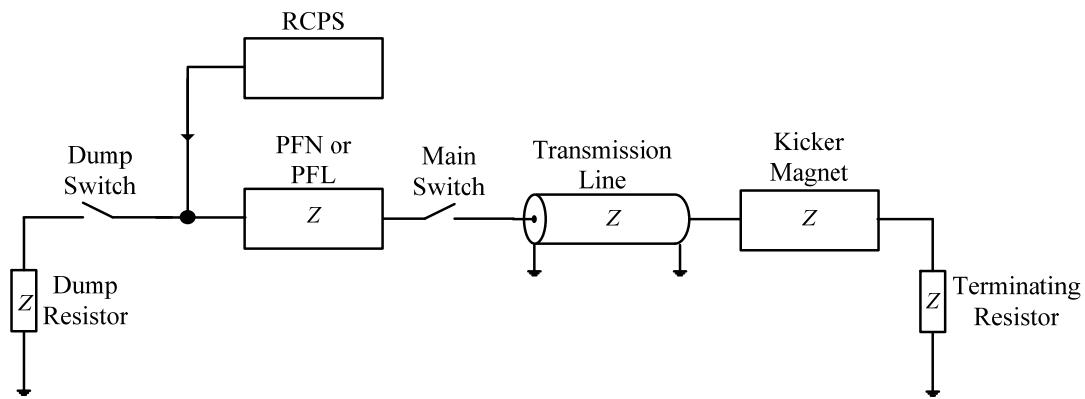

**Fig. 4:** Simplified schematic of a kicker system

### 4.2 Pulse forming circuit: general case

Figure 5 shows a simplified schematic of a pulse-forming circuit: the switch is initially open and the coaxial cable (PFL) is pre-charged, through the large valued resistor or inductor, to a voltage $V$.

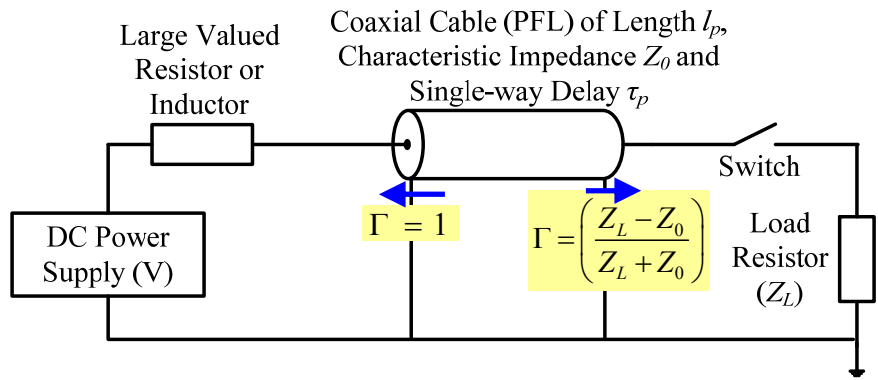

**Fig. 5:** Simplified schematic of a pulse-forming circuit

At time $t = 0$, when the ideal switch closes, the load voltage is given by Eq. (1):

$$V_L = \left(\frac{Z_L}{Z_0 + Z_L}\right) V = \alpha V, \qquad (1)$$

where

$$\alpha = \left(\frac{Z_L}{Z_0 + Z_L}\right);$$

$V_L$      is the load voltage (V);

$V$      is the initial voltage to which the PFL is charged (V);

$Z_L$      is the load impedance (Ω);

$Z_0$      is the characteristic impedance of the PFL (Ω).

Figure 6 shows the lattice diagram for the general case (impedances not necessarily matched) for the voltage on the PFL. At time $t = 0$, when the switch closes, a voltage pulse of '$(\alpha - 1)V$' propagates from the load end of the PFL towards the charging end. At the charging end of the PFL the reflection coefficient ($\Gamma$) is +1 and hence a voltage of '$(\alpha - 1)V$' is reflected back towards the load end of the PFL. At the load end of the PFL the reflection coefficient is given by Eq. (2):

$$\Gamma = \left(\frac{Z_L - Z_0}{Z_L + Z_0}\right) = \beta, \qquad (2)$$

and hence a voltage of '$\beta(\alpha - 1)V$' is reflected back towards the charging end of the PFL, etc.

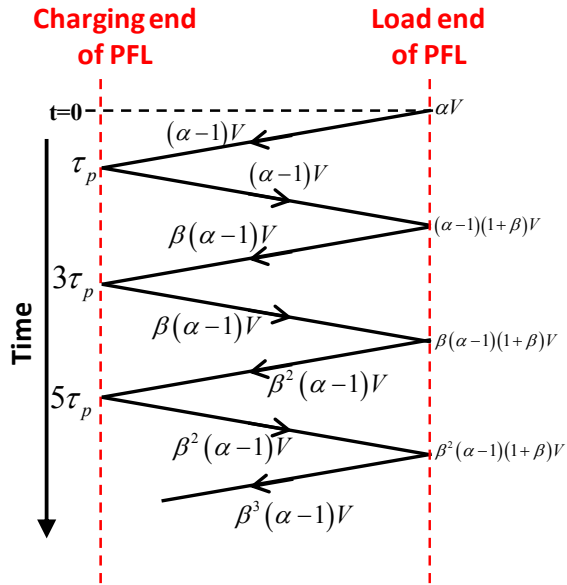

**Fig. 6:** Lattice diagram for the pulse forming circuit of Fig. 5: general case

Impedances need to be matched to avoid reflections, i.e., from Eq. (2), when $Z_L = Z_0 \rightarrow \beta = 0$. When the impedances are matched:

- PFN/PFL is charged to a voltage $V$ by the RCPS;
- The Main Switch (MS) closes (Fig. 4) and, for a matched system, a pulse of magnitude $V/2$ is launched, through the transmission line, towards the kicker magnet. A voltage pulse of magnitude $-V/2$ propagates from the load end of the PFN/PFL towards the charging end;
- Once the current pulse reaches the (matched) terminating resistor full-field has been established in the kicker magnet (Section 4.3.3);
- The length of the pulse in the magnet can be controlled in length, between 0 and $2\tau_p$, by adjusting the timing of the Dump Switch (DS) (Fig. 4) relative to the MS.

Note: if the magnet termination is a short-circuit, the magnet current is doubled but the required 'fill-time' of the magnet is doubled too (Section 4.3.6). In this case the DS may be an inverse diode: the inverse diode 'automatically' conducts when the PFN voltage reverses, at the charging end of the PFL/PFN, but there is no control over the pulse-length in the magnet.

### 4.3 Kicker magnet

#### 4.3.1 History

Figure 7 shows a 1970's 'plunging' kicker magnet which was hydraulically operated [2]: the aperture was too small for the kicker to be in the beam-line during circulating beam. Developments leading to higher current pulses permitted larger apertures: kicker magnets developed later at CERN were not hydraulically operated.

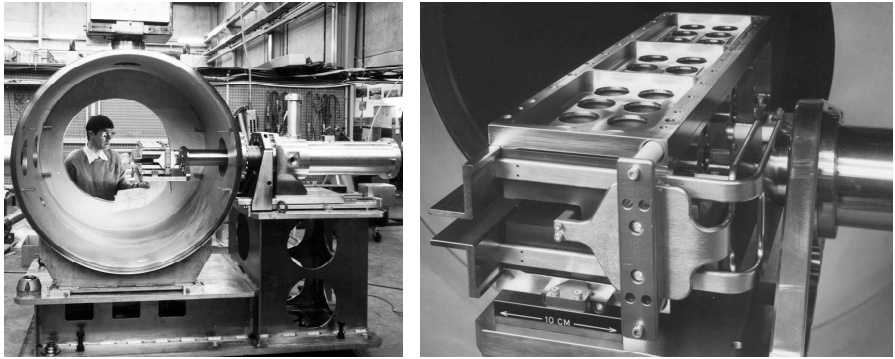

**Fig. 7:** 1970's plunging kicker magnet for ejection from the CERN Antiproton Accumulator (AA)

#### 4.3.2 Overview

Figure 8 shows a cross-section of a typical C-core kicker magnet. Fast kicker magnets are generally ferrite-loaded transmission line type magnets with a rectangular-shaped aperture of dimensions $H_{ap}$ by $V_{ap}$ (Fig. 8).

The flux density in the aperture ($B_y$) of the kicker is given by Eq. (3):

$$B_y \cong \mu_0 \left( \frac{N \cdot I}{V_{ap}} \right), \tag{3}$$

where

- $\mu_0$     is permeability of free space ($4\pi \times 10^{-7}$ H/m);
- $N$     is the number of turns (usually 1 for a kicker magnet);
- $I$     is magnet current (A);
- $V_{ap}$     is the distance between the inner edges of the 'legs' of the ferrite (m).

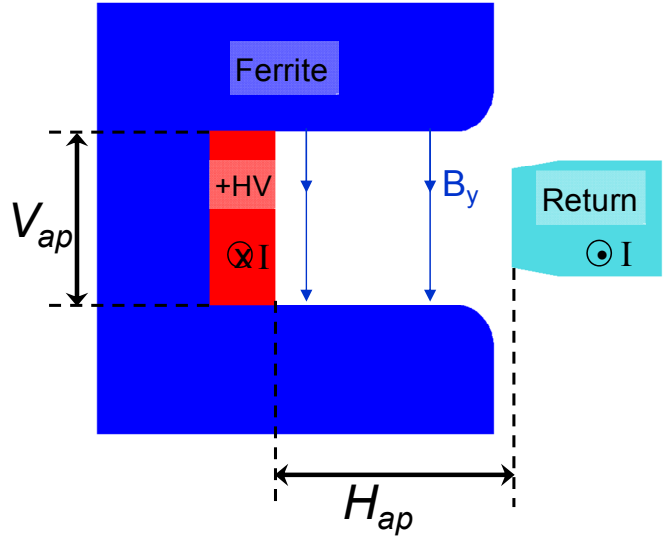

**Fig. 8:** Cross-section of a typical C-core kicker magnet

Skin effect and proximity effect result in current flow on the inside surface of both the High Voltage (HV) and return conductors. Hence inductance is given by Eq. (4):

$$L_{m/m} \cong \mu_0 \left( \frac{N^2 \cdot H_{ap}}{V_{ap}} \right), \qquad (4)$$

where

- $H_{ap}$     is the distance between the inner edges of the HV and return conductors (m);
- $L_{m/m}$     is inductance per metre length of the kicker magnet (H/m).

Since kicker magnets generally need to be fast they usually have only single-turn coils: multi-turn coils are used only for slower, lumped-inductance kicker magnets.

The deflection of a charged particle beam in a magnetic field and an electric field is given by Eq. (2) and Eq. (3), respectively, of *Injection and extraction magnets: septa* in the proceedings of this CAS. For a transmission line kicker magnet, where ferrite C-cores are sandwiched between HV capacitance plates (see Section 4.3.3.2), end effects result in an effective length of each end-cell approximately $(V_{ap}/4)$ greater than its physical length [3]. Thus the effective length of the kicker magnet is increased by an amount approximately equal to $(V_{ap}/2)$, in comparison to the physical length of the aperture.

### 4.3.3 Design options for kicker magnets

Design options for kicker magnets include [2]

– Type: 'lumped-inductance' or 'transmission-line' (with specific characteristic impedance ($Z$))?
– Machine vacuum: install in or external to machine vacuum?
– Aperture: window-frame, closed C-core or open C-core?
– Termination: matched impedance or short-circuited?

#### 4.3.3.1 Lumped-inductance kicker magnet

Although a lumped-type magnet has a simple structure, in most cases it cannot be applied to a fast kicker system because of its impedance mismatch and its slow response [4]. The lumped-inductance kicker is generally useable only when a rise-time above a few hundreds of nanoseconds is required. The lumped-inductance kicker either has a resistor in series with the kicker magnet input or else the resistor is omitted. In both cases the kicker magnet only sees (bipolar) voltage during pulse rise and fall. With a short-circuit termination, magnet current is doubled for a given PFN/PFL voltage and system characteristic impedance.

For a magnet inductance $L_m$ in series with a parasitic inductance $L_s$, combined with a pulse generator of impedance $Z$, the rise of the magnet current is exponential with a time constant $t_{cl}$ given by Eq. (5):

$$t_{cl} = \left( \frac{L_m + L_s}{Z} \right) . \tag{5}$$

For a short-circuit magnet the 5% to 95% rise-time of the current is as long as three time-constants. To help to overcome the long rise-time a capacitor can be connected in parallel with, at the entrance of, the lumped-inductance magnet, but this can provoke some overshoot.

#### 4.3.3.2 Transmission-line kicker magnet

To overcome the long rise-time, the first transmission-line kicker magnet was developed at CERN in the early 1960s [4]. A transmission-line magnet consists of few to many 'cells' to approximate a broadband coaxial cable (Fig. 9). Ferrite C-cores are sandwiched between HV capacitance plates: plates connected to ground are interleaved between the HV plates. The HV and ground plates form a capacitor to ground (Fig. 9). One C-core, together with its ground and HV capacitance plates, is termed a cell. Each cell conceptually begins and ends in the middle of the HV capacitance plates. The 'fill-time' of the magnet ($\tau_m$) is the delay required for the pulse to travel through the '$n$' magnet cells.

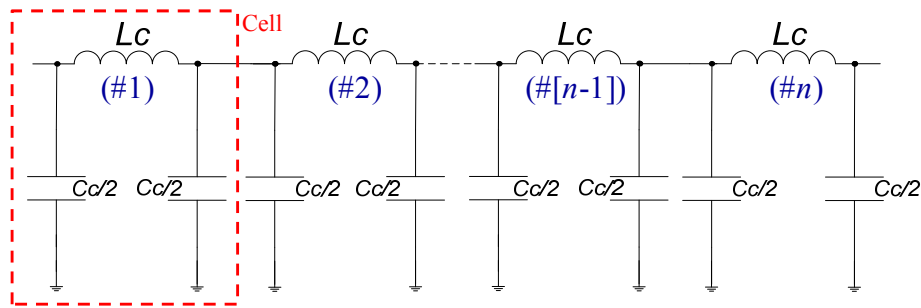

**Fig. 9:** Simplified equivalent electric circuit of a transmission-line kicker magnet

The characteristic impedance ($Z$) of the kicker magnet, which is matched to the impedance of the generator to minimize reflections [5, 6] (see Section 4.2), is given by Eq. (6):

$$Z = \sqrt{\frac{L_c}{C_c}}, \tag{6}$$

where

- $L_c$      is the inductance of a cell of the kicker magnet (H),
- $C_c$      is the capacitance of a cell of the kicker magnet (F).

The fill-time of the kicker magnet is given by Eq. (7):

$$\tau_m = n\sqrt{L_c \cdot C_c} = n\left(\frac{L_c}{Z}\right) = \left(\frac{L_m}{Z}\right), \tag{7}$$

where

neglecting end effects, $L_m = n(L_c)$,

- $L_m$      is the inductance of the kicker magnet (H).

For a kicker magnet terminated with a matched resistor, field rise-time starts with the beginning of the voltage pulse at the entrance of the kicker magnet and ends with the end of the same pulse at the exit of the magnet. Flux is given by the time integral of the difference between the voltage at the magnet entrance and exit [Eq. (8)]:

$$\Phi = \int (V_{in} - V_{out}) \, dt, \tag{8}$$

where

- $\Phi$      is flux (V·s),
- $V_{in}$      is the voltage at the entrance to the kicker magnet (V),
- $V_{out}$      is the voltage at the exit of the kicker magnet (V).

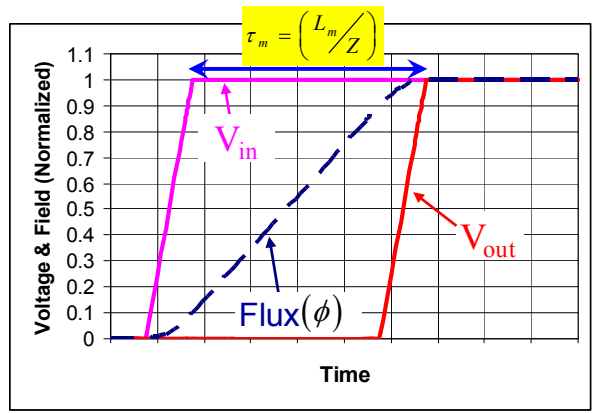

**Fig. 10:** Flux in an 'ideal' transmission-line kicker magnet

The flux builds up until the end of the voltage rise at the exit of the kicker magnet (Fig. 10), thus, for an ideal transmission-line kicker magnet, flux rise-time is given by the sum of the voltage rise-time and the magnet fill-time. Therefore it is important that the pulse not degrade while travelling through the kicker magnet. Hence the magnet cut-off frequency is a key parameter, especially with field rise-times below ~100 ns. Cut-off frequency ($f_c$) depends on series inductance ($L_c s$) associated with the cell capacitor ($C_c$):

$$f_c = \frac{1}{\pi \cdot \sqrt{(L_c + 4 L_c s) \cdot C_c}} = \frac{Z}{\pi \cdot \sqrt{(L_c + 4 L_c s) \cdot L_c}} \quad . \tag{9}$$

Thus, to achieve a high cut-off frequency, $L_c s$ should be kept as low as possible and the cell size small. However, cells cannot be too small because adequate distance is required between HV and ground capacitance plates to avoid voltage breakdown. In addition many very small cells would significantly increase the cost and complexity of the kicker magnet.

Figure 11 shows the results of a low-voltage measurement on each of the HV capacitance plates of a transmission-line kicker magnet. The fast rise-time of the input voltage pulse, used for the measurements, contains frequency components above the cut-off frequency of the cells. Thus there is an increase in the rise-time of the voltage pulse between the entrance HV capacitance plate and the second HV capacitance plate; in addition there is significant ripple on the pulses.

The choice of the characteristic impedance, for a transmission-line magnet, depends upon the requested field rise-time and the available length for the kicker magnet. In general the highest impedance, up to the impedance of commercially available coaxial cable (50 Ω), is used while still respecting the available length for the kicker magnet and chosen level of PFN voltage [2]. The higher the characteristic impedance, the higher the magnet cut-off frequency [Eq. (9)]: a high cut-off frequency reduces the field ripple.

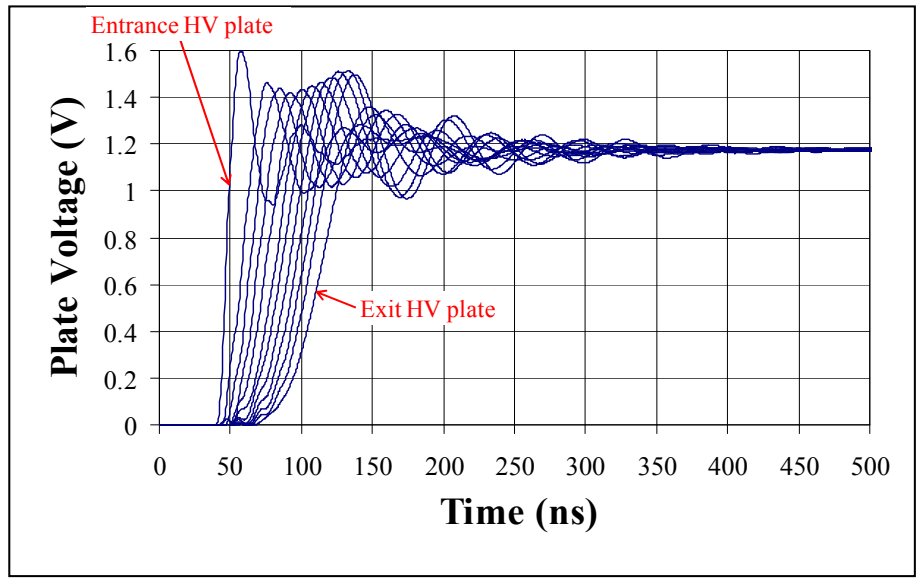

**Fig. 11:** Low voltage measurement on each of twelve HV capacitance plates of an eleven-cell transmission-line kicker magnet

A vacuum dielectric transmission-line kicker magnet can require a very large area of metal plates to form the capacitance, especially for a low-impedance system. Since the minimum separation of the capacitor plates is determined by practical considerations, such as voltage breakdown and cost, the capacitance is determined by the area of the plates: however, Eq. 6 shows that the capacitance (area) must increase by a factor of 4 for a decrease in impedance by a factor of 2 (neglecting edge effects). It has been shown that it is feasible to design an 'hybrid' kicker magnet, which is an effective transmission kicker magnet with a fast rise-time using only 3 to 5 large cells [7], but which requires large capacitance values. However, it is impractical for lower impedance magnets or for magnets with larger cell inductances to use vacuum dielectric capacitors. Hence, with the space limitation at many accelerator facilities, the application of higher permittivity dielectric capacitor media is necessary [6, 8].

*4.3.4 Machine vacuum*

The minimum-sized aperture for a kicker magnet can be achieved if the magnet is in vacuum: if the magnet is outside vacuum a chamber must be inserted in the kicker aperture thus increasing the dimensions of the aperture. The minimum value of both $H_{ap}$ and $V_{ap}$ (Fig. 8) are determined by beam parameters. Equation (4) shows that magnet inductance is proportional to $H_{ap}$ and inversely proportional to $V_{ap}$. However, if $V_{ap}$ is increased, although the magnet inductance is reduced, this is at the expense of increased current to obtain a given flux density (Eq. 3): increased current also requires, for given impedance, increased PFN voltage and thus increased insulation requirements.

Advantages of putting a transmission line kicker magnet in the machine vacuum are

– Aperture dimensions are minimized, therefore the number of magnets and/or voltage and current are minimized for a given ∫ B.dl and rise-time.

– Machine vacuum is a reliable dielectric (70 kV/cm is OK [2]) and generally 'recovers' after a flashover, whereas a solid dielectric, outside vacuum, may not recover.

Disadvantages of putting a transmission line kicker magnet in the machine vacuum are

– The kicker magnet is costly and time consuming to construct (all parts must be appropriately cleaned and handled, a vacuum tank is required, suitable pumps are needed, the kicker magnet must be baked-out and the design must allow for the thermal expansion during bake-out, etc.).

– In the event of failure of something inside the vacuum tank the vacuum must be broken to repair the fault. Afterwards pumping, bake-out, and HV conditioning are required — a time-consuming process.

The beam coupling impedance may be an issue irrespective of whether the magnet is in vacuum. Hence a beam screen (see Section 4.5), in the kicker aperture, may be required which will require an increase in aperture dimensions in any case.

*4.3.5 Kicker magnet aperture*

Normally a magnetic circuit is used which contains magnetic material [2, 4]: without magnetic material the effective value of $V_{ap}$ (Fig. 8) is greatly increased, therefore requiring more current to achieve the required field (Eq. 3). In addition, magnetic material improves field uniformity. Nickel-zinc (NiZn) ferrite, with $\mu_r \approx 1000$, is usually used [2, 9]. If the ferrite is in machine vacuum (Section 4.3.4) the out-gassing properties must be acceptable: in addition, the density of the ferrite must be such that its water absorbent characteristics are not too high and no undesirable lubricants must be used during the grinding process [9]. NiZn ferrite has the following properties:

- field rise can track current rise to within ~1 ns [2],
- low remnant field,
- low out-gassing rate, after bake-out.

Figure 12 shows a full-aperture C-core magnet and a window-frame magnet. The C-core magnet shown in Fig. 8 has its aperture closed by the return conductor. The C-core magnet on the Left Hand Side (LHS) of Fig. 12 has the return conductor behind the yoke: this is for beam gymnastic reasons and has the effect of increasing the effective width ($H_{ap(eff)}$) of the magnet aperture, of given dimensions, from $H_{ap}$ to a value given by Eq. (10) [10, 11]:

$$H_{ap(eff)} \approx \left(H_{ap} + V_{ap}/2\right) . \tag{10}$$

Similarly the inductance of a cell of the full-aperture kicker magnet is increased by a factor of $\left(H_{ap(eff)}/H_{ap}\right)$ with respect to the return conductor closing the aperture of the kicker magnet.

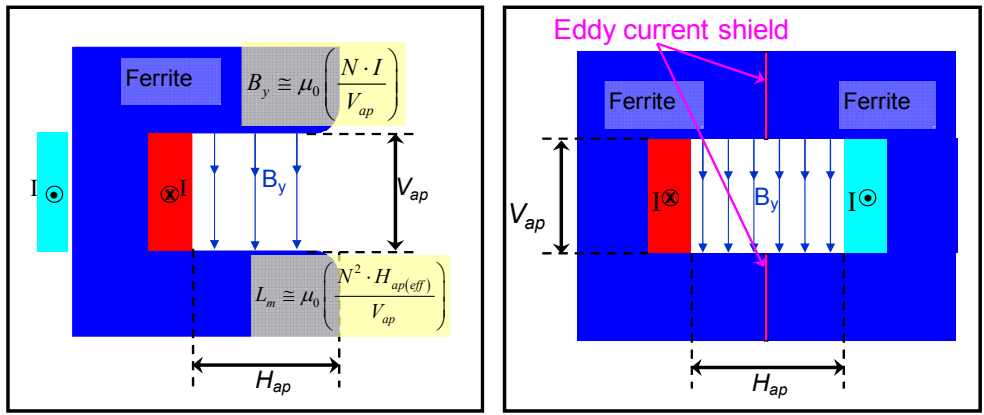

**Fig. 12:** Full-aperture C-core magnet (LHS) and window frame magnet (RHS)

To reduce magnet fill-time by a factor of two, FNAL and KEK use a window-frame topology (Right Hand Side (RHS) of Fig. 12): it can be considered as two symmetrical C-magnets energized independently. However, the window-frame magnet requires two generators to achieve the reduced magnet fill-time [4]. It is necessary to include eddy current 'shields' between the two ferrite C-cores to reduce beam coupling impedance (see Section 4.5).

C-core kicker magnets are generally used at CERN.

### *4.3.6 Kicker magnet termination*

When space along the beam line is at a premium, a short-circuit termination has the advantage over a matched resistive termination of doubling kick (for a given system impedance) [2]: in addition, a short-circuit termination reduces the time during which the kicker magnet is exposed to high voltage. However, disadvantages include:

- fill-time of the kicker magnet is doubled: to establish full field in the magnet, the current/voltage pulse must propagate from the entrance to the exit of the magnet and then reflect back to the input;
- the kicker magnet experiences voltage of both polarities: the incident voltage pulse is inverted at the short-circuit termination (reflection coefficient is −1);

- if the DS is used to control pulse length it must be bidirectional (unidirectional dump-switch, acting as an inverse diode, is suitable for a fixed length pulse);
- beam coupling impedance can be affected [12] (owing to resonances, below magnet cut-off frequency, with kicker circuitry).

**4.4 Kicker magnet design tools**

Circuit simulation and finite element codes greatly assist the goal of obtaining high performance kicker systems. Simulation of circuits which include almost all known parasitic elements and non-linearities is now possible [4].

2D and 3D finite element codes now include AC and transient analysis with eddy currents. These tools are used for kicker magnet design to predict magnetic field, cell inductance, and electric field [13–15]. In order to obtain realistic predictions for inductance and magnetic field distribution, in the aperture of the kicker magnet, it is necessary that the skin effect and proximity effect, in the HV and 'ground' (return) conductors (Fig. 13), be properly accounted for: this requires an AC or transient analysis to be carried out. Codes such as Opera2D and Opera3D [16] are also used to study the shape of the ferrite and conductors and thus optimize the field homogeneity [14, 15]; the current distribution is calculated by the code. The LHS of Fig. 13 shows a transmission-line kicker magnet used to deflect beam horizontally. The magnet was modelled using Opera2D to predict magnetic and electric field distribution. The predictions were post-processed and the deflection calculated; the RHS of Fig. 13 shows a plot of deflection uniformity for the optimized geometry of ferrite and return conductor [14, 15].

The penetration of the pulsed field through a beam screen (Section 4.5), in a magnet aperture, can also be accurately calculated and the frequency dependence of magnet inductance, due to eddy currents in the screen, predicted [17]. An equivalent circuit can then be fitted to the resulting predictions to account for the frequency dependence in an analog circuit simulation [18].

An AC analysis can also be used to predict the frequency dependence due to eddy currents in PFN coils [18, 19] (see Section 4.6). An equivalent circuit can be fitted to the resulting predictions to account for the frequency dependence in an analog circuit simulation.

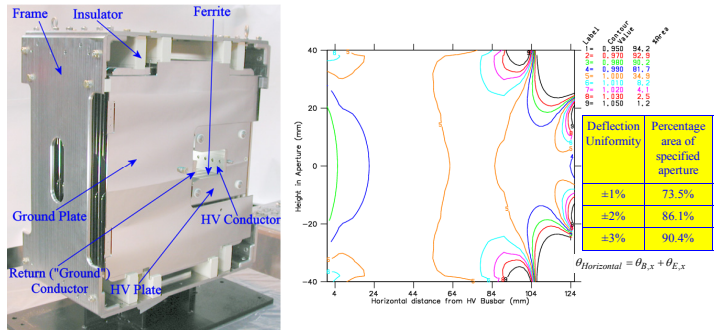

**Fig. 13**: Transmission-line kicker magnet and corresponding deflection uniformity plot

In addition capacitance to ground of an HV plate, in a kicker magnet, is influenced by insulators and nearby ground planes:

- ground plate,
- magnet frame,
- return conductor.

Boundary element software such as *Coulomb*, a 3D code from Integrated Engineering Software [20], can be used to accurately predict capacitance of a cell of a kicker magnet. The predictions resulting from *Coulomb* have been carefully checked against measurements and found to be in good agreement [3].

### 4.5 Beam-coupling impedance

High-intensity accelerators are very sensitive to longitudinal and transverse beam-coupling impedance. Kicker magnets, with their ferrite yoke, can result in considerable beam-coupling impedance [21, 22]. The beam coupling also induces heating of the ferrite yoke [23, 24] sometimes beyond the Curie temperature. In extreme cases it can affect the generator as well [25].

Some beam-coupling high-impedance resonances are attributable to the image path of the beam not being continuous, for example, a gap between the end of the kicker magnet and the vacuum tank. In this case these high-impedance resonances can be eliminated by inserting transition pieces inside the tank to electrically connect the tank flange to the kicker magnet at each end [23, 26, 27]. When several kicker magnets are installed in a common tank, the transition between magnets is carried out in a similar way [4].

Beam-coupling impedance of the ferrite yoke is addressed by providing a beam screen in the magnet aperture [12, 17, 21, 22, 24, 28–30]. The screen conducts the beam image current but at the same time must not significantly attenuate the pulsed field; thus eddy-current loops must be avoided.

The LHS of Fig. 14 shows beam-coupling impedance reduction techniques currently being implemented for the MKE kickers [21], used for extraction from the CERN SPS ring. Silver combs are printed directly on the ferrite, by serigraphy, and connected to the high-voltage plates. The serigraphy does not reduce the available aperture, and the high permittivity of the ferrite enhances the capacitive coupling between the serigraphy: the capacitive coupling provides a continuous path for the beam image current while preventing eddy-current loops. This solution is less attractive for magnets with short cells since the capacitive coupling is reduced.

The RHS of Fig. 14 shows beam-coupling impedance reduction techniques implemented for the MKI kickers, used for injection into the CERN LHC rings. A 3 m long ceramic tube is manufactured with 24 slots on its inside diameter [17]. Carefully radiused conductors are inserted into the slots: the ceramic tube acts as a support and in addition provides insulation to the ferrite and high-voltage plates. Capacitive coupling is made at one end between the beam screen conductors and an outside metallization which is connected to ground (Fig. 14): the capacitive coupling provides a continuous path for the beam image current while preventing eddy-current loops. 3D electromagnetic simulations have been carried out to optimize the geometry of the beam screen [29].

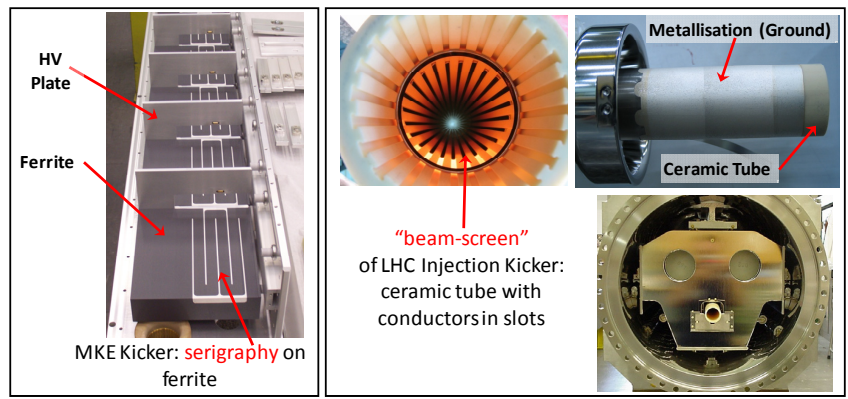

**Fig. 14:** Beam-coupling impedance reduction techniques for the MKE (LHS) and MKI (RHS) kickers

Figure 15 shows longitudinal beam-coupling impedance for several systems: unless mentioned otherwise the impedance is derived from measurements.

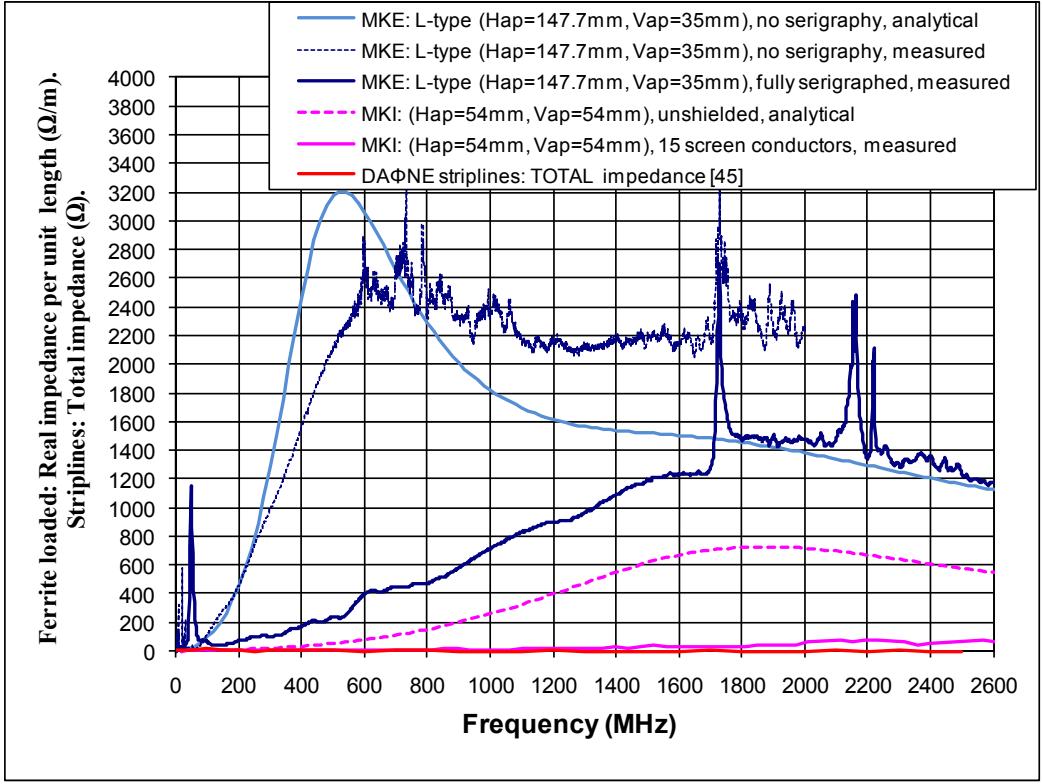

**Fig. 15:** Longitudinal beam-coupling impedance for several systems

Longitudinal beam-coupling impedance is significantly reduced by

- serigraphy of ferrites (painted stripes): there is negligible loss of aperture;
- beam screen conductors within the kicker magnet aperture: however this can result in approximately 15 mm loss of aperture and is thus usually not applicable as a retrofit to an existing kicker magnet;
- striplines instead of a ferrite loaded magnet: this is not feasible for obtaining a large kick in a limited length.

### 4.6 Pulse-forming line/network

The simplest configuration for a pulse-forming circuit is a PFL (coaxial cable) charged to twice the required pulse voltage. The PFL gives fast and ripple-free pulses, but low attenuation is essential, especially with longer pulses, to keep droop and 'cable tail' within specification [2]. Attenuation is adversely affected by the use of semiconductor layers to improve voltage rating [2]. Hence, for PFL voltages above 50 kV, SF6 pressurized polyethylene (PE) tape cables are used. PFL becomes costly, bulky, and the droop becomes significant (e.g. ~1%) for pulses exceeding about 3 μs duration.

Where low droop and long pulses are required a PFN is used: a PFN is an artificial coaxial cable made of lumped elements. The SPS extraction PFNs at CERN [31, 32], which are approximately 30 years old, have 17 cells, connected in series in a single line (Fig. 16, RHS), which are individually

'adjustable': these PFNs have 'corners' (Fig. 16, LHS), therefore mutual inductance between cell inductances is not well defined and can result in ripple on the pulse. Adjusting the pulse flat top is difficult and time-consuming.

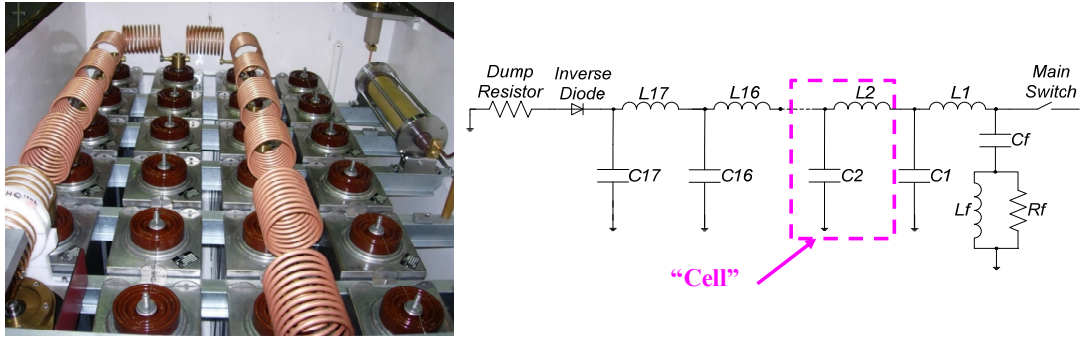

**Fig. 16:** The SPS extraction PFNs at CERN have 17 cells which are individually 'adjustable'

The preservation of the transverse emittance of the proton beam at injection into the LHC is crucial for luminosity performance. The transfer and injection process is important in this respect, and injection offsets are a well-known source of error [32]. To limit the beam emittance blow-up due to injection oscillations, the reflections and the flat top ripple of the field pulse must be lower than ±0.5%, a very demanding requirement [8].

Each of the two LHC injection kicker systems has four 5 Ω PFNs [33]. Each PFN consists of two lumped element delay lines, each of 10 Ω, connected in parallel [33]. Each 10 Ω line consists of 26 central cells plus two end-cells. A cell consists of a series inductor, a damping resistor connected in parallel, and a capacitor connected to ground (Fig. 17, LHS). The inductors are part of a single continuous coil, 4.356 m long, with 198 turns and a pitch of 22 mm [33]. The central cell inductors are made of seven turns each. The nominal MS and DS end-cell inductors have nine turns and five turns respectively, but are built with one extra turn to allow some adjustment to compensate for end-effects. The 26 central cells of the coils are not adjustable and therefore are defined with high precision: the coil conductor is a copper tube wound on a rigid fibreglass coil former. Both delay lines are mounted in a rectangular tank (Fig. 17, RHS), with mild steel walls, that is filled with insulating silicone fluid. Each line is surrounded by a 3 mm thick, Ω-shaped, aluminium shield, which has an inner radius of 140 mm. Two thyratron switches, a MS and a DS, are connected to the PFN.

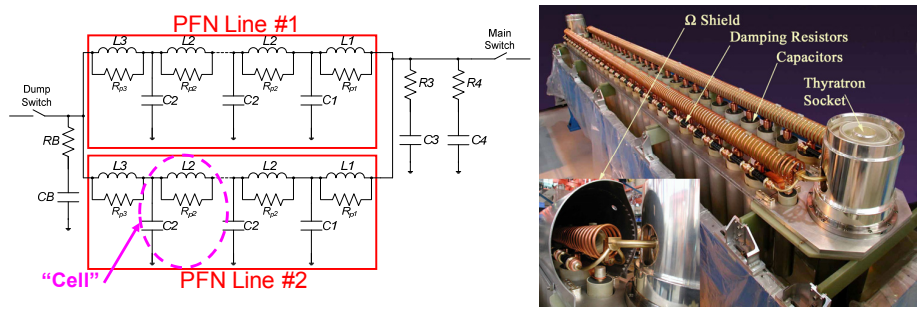

**Fig. 17:** Each LHC injection PFN consists of two parallel 10 Ω lines with precision wound coils

Opera2D simulations have been carried out to assess the frequency dependence of inductance and resistance of a coil of the LHC injection PFN [19]. Figure 18 shows the predicted current distribution in the PFN coil at frequencies of 0.1 Hz, 200 Hz, 1 kHz and 40 kHz: the nominal DC current density is 27.3 kA/cm$^2$, for a total current of 6 kA.

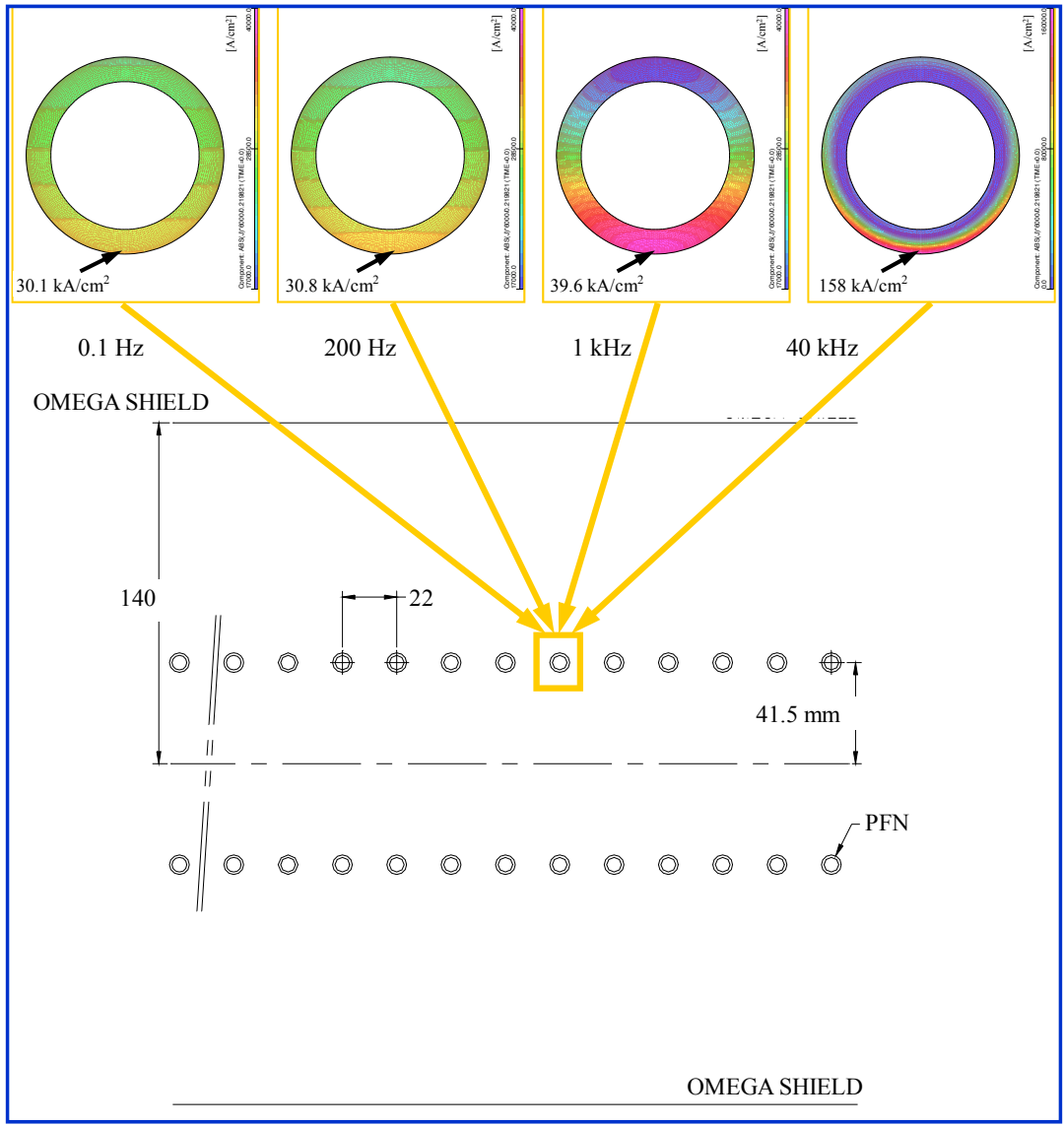

**Fig. 18:** Opera 2D prediction for current distribution in LHC injection PFN for a current of 6 kA

Figure 19 shows the predicted inductance of a 7-turn cell versus frequency for a mean radius of the coil of 41.5 mm. The 'Grover limits' refer to values calculated from equations [19]. The reduction in inductance as frequency is increased from DC to a few hundred Hertz is mainly due to screen shielding. The reaction field from the eddy currents induced in the Omega shield reduces the flux density along the axis of the coil from 0.343 T near DC to 0.315 T at a few hundred Hertz, for a current of 6 kA. As the frequency is increased beyond a few hundred Hertz the inductance decreases, mainly due to skin effect and proximity effect within the coil. Conduction losses along the coil result in droop of the pulse of approximately 0.5% in the kicker magnet. PSpice simulations, subsequently confirmed by measurements, show that conduction losses in the PFN coil can be compensated for by grading the PFN capacitor values linearly from the MS end to the DS end [33].

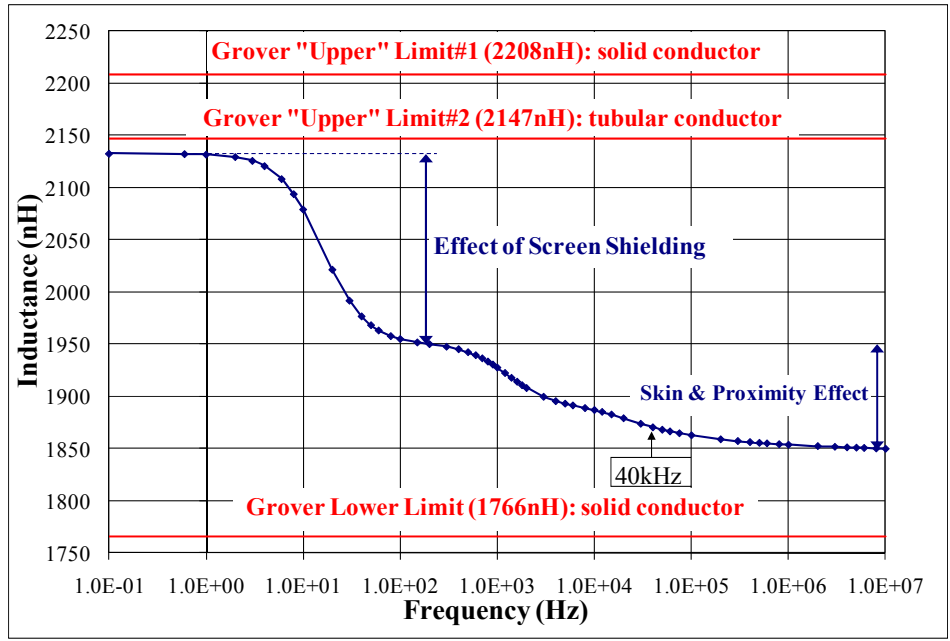

**Fig. 19:** Opera 2D prediction for inductance of a 7-turn PFN cell, versus frequency, for a mean radius of the coil of 41.5 mm

### 4.7 Power switches

#### *4.7.1 Thyratrons*

Despite the advances in high-power solid-state switches, deuterium thyratrons are still widely used as the power switch in kicker systems. The deuterium thyratron is a high peak power electrical switch which uses deuterium gas as the switching medium. The switching action is achieved by a transfer from the insulating properties of neutral gas to the conducting properties of ionized gas [34]. Voltage breakdown in the deuterium-filled gap is initiated by free charges (electrons and ions) crossing the gap under the influence of an electric field. If sufficient energy is available, gas molecules are ionized producing more free charges. The positive ions are accelerated towards the lower potential electrode and cause the release of secondary electrons. Under the right circumstances, the processes become self-sustaining and voltage breakdown occurs [34].

Thyratron commutation is achieved by introducing plasma into the grid/anode region via slots in the grid structure. The plasma is created in the cathode/grid region by a fast-rising trigger pulse applied to the grid(s), which then diffuses to the grid slots where it comes under the influence of the anode field [34].

Three-gap thyratrons can hold-off 80 kV and switch 6 kA of current with a 30 ns rise-time (10% to 90%) [~150k A/µs]. However, special care must be taken with the high-power thyratron:

– Coaxial housings are used to achieve low inductance but there must be adequate insulation to the housing.

– Current flow in the connections around the thyratron should be symmetrical. An asymmetrical magnetic field can impose forces on the internal plasma preventing uniform current density which may affect performance and lifetime [34].

- The appropriate thyratron must be selected (e.g., suitable rating for peak current and average current) and properly applied for the anticipated short-circuit and fault conditions.

- The thyratron must be adequately cooled to ensure that the maximum specified envelope temperature is not be exceeded.

- Erratic turn-on (turn-on without a trigger being applied) can result in operational problems: erratic turn-on is reduced significantly by 'fast' (~ms) charging of the PFN/PFL immediately before the kicker magnet is required to be pulsed.

- The reservoir voltage of the thyratron should be optimized. The gas density in the grid/anode gap must be maximized, to minimize current rise-time and switching losses and maximize thyratron lifetime, consistent with achieving a low rate of erratic turn-on.

In order to reduce as much as possible the number of erratic turn-ons, Resonant Charging Power Supplies (RCPS) are used to charge PFNs for kicker magnets [2, 35]. The number of erratic turns-on of a thyratron switch is dependent upon several variables which include the time period for which there is a high voltage across the thyratron: thus, in order to minimize the number of erratics, it is advantageous to minimize this time period. The fast RCPS developed for the LHC injection kicker systems charges two PFNs to 54 kV in approximately 800 μs: approximately 600 μs later the PFNs can be discharged into the kicker magnets [36]. An RCPS can also allow the reservoir voltage to be increased, provided that the rate of erratic turn-on is still acceptable, and therefore help to extend thyratron lifetime.

To achieve significantly improved performance and lifetime a thyratron can be double pulsed to turn it on: the first pulse pre-ionizes the cathode grid space which prepares the cathode region for conducting the main current pulse. The second pulse, which should be delayed by at least 500 ns, provides a fast rising voltage to ensure precise triggering (1–2 ns) of the thyratron [34].

*4.7.2  Power semiconductor switches*

In some applications thyratron switches cannot be used, e.g., for the dump (abort) kickers in the LHC where no self-firing is allowed [37]. For the LHC system, acceleration of the beam to the required energy and subsequent physics may require more than 10 hours: the generators of the LHC dump kickers must be charged to the correct voltage, which must track the beam energy, and be ready to operate at any time in the LHC cycle. In this application high-power semiconductor switches are used. The semiconductor switches also allow a wide dynamic range of operation [37]. Maintenance is significantly reduced with a semiconductor switch, in comparison with a thyratron. However, high-power semiconductors have a maximum rate-of-rise of current which is significantly less than a thyratron: the current rise-time for a high-power semiconductor switch is typically ~1 μs.

Semiconductors may be influenced by ionizing radiation and neutron flow. Depending on the type of radiation, the component type and its working conditions, the radiation effects can be cumulative with relatively slow deterioration of semiconductor performance and/or sudden malfunction or failure: the latter is called 'Single Event Effects' (SEE). The cumulative effects, which result from Total Ionizing Dose (TID) and displacement damage, are responsible for the modification of component parameters such as leakage current, bipolar transistor gain, opto-coupler efficiency, MOS threshold voltage, voltage reference value, etc. On the other hand, a SEE provokes sudden malfunctions which, in the case of HV power semiconductors, usually lead to component failure: this is known as Single Event Burnout (SEB). For a given radiation dose the SEB rate is strongly dependent on the applied voltage: there is a very steep increase of the failure rate when the applied voltage is higher than a certain percentage of the components' rated voltage. The general recommendation for high voltage components, to maintain a reasonably low SEB rate, is to apply a maximum DC voltage of approximately 50% of the semiconductor rating. For the LHC abort kickers the power semiconductors can be affected by radiation and neutron flow from the LHC [38].

Each semiconductor switch module for the LHC abort kickers consists of ten series Gate Turn Off (GTO) thyristors (Fig. 20) that have been optimized for turn-on. The maximum forward voltage rating of each GTO is 4.5 kV and the switch module is operated over a voltage range from 2.2 kV (450 GeV) to 30 kV (7 TeV) [37]. The current switched is in the range from 1.3 kA to 18.5 kA: the maximum rate of rise of current is 18 kA/μs, which corresponds to approximately 1/8th of the capability of a thyratron. However, recent tests on similar GTOs, with a high-current gate drive, have demonstrated a capability of more than 32 kA/μs.

Ultra-fast kicker systems generally use either fast high-voltage MOSFETs [39–45] or a Fast Ionization Dynistor (FID) [45–47]: the FID is also sometimes called a Fast Ionization Device.

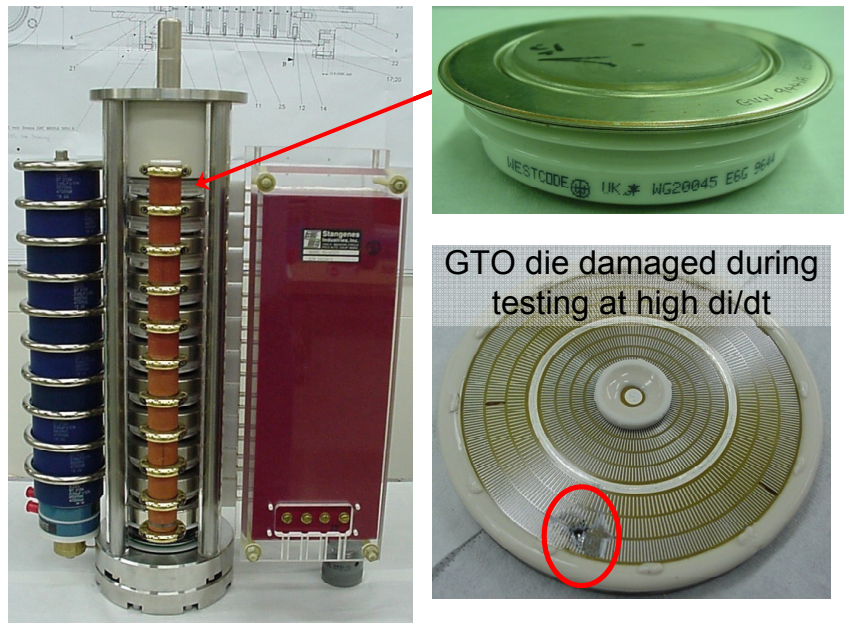

**Fig. 20:** Stack of-high power GTOs (LHS) and individual GTOs (RHS)

### 4.8 Ultra-fast kickers

Where sub-nanosecond jitter is required, only semiconductor switches can be used. A case in point is the tail-clipper kicker for the CLIC Test Facility 3 (CTF3) [48] at CERN: the tail-clipper must have a fast field rise-time, of 5 ns or less, to minimize uncontrolled beam loss. For this application there are eight pulse generators: each is composed of a 50 Ω PFL, a fast semiconductor MOSFET switch, 50 Ω stripline plates (no magnetic material) and a matched terminating resistor [43, 44]. The deflection of the electron beam makes use of both the electric and magnetic field in the stripline plates: in order that the effects of the electric and magnetic field do not cancel, the striplines are fed from the beam exit end of the plates (Fig. 21).

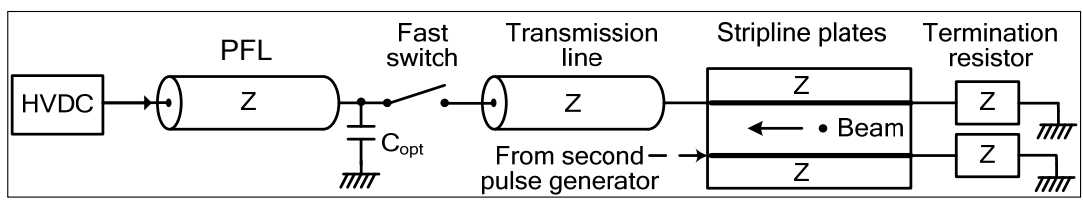

**Fig. 21:** Schematic of CTF3 tail-clipper kicker

The overall length of the stripline assembly is mechanically sub-divided into four sections of equal length of 380 mm; a pair of stripline plates is energized a time delay of 1.27 ns (0.38 m/c) after the previous section to minimize the overall apparent rise-time of the kick [43]. Each set of stripline plates is driven by two fast switches, one connected to a positively charged PFL and the other connected to a negatively charged PFL (Fig. 21). Figure 22 shows a measured current pulse from a MOSFET switch: the 56 A of current has a 10% to 90% rise-time of 2.5 ns. A novel design of gate driver results in an overall 3-sigma jitter of less than ±300 ps for the modulator system [43].

A FID was considered for use for the CTF3 tail-clipper but a suitable device was not delivered in time: thus the FID is currently being evaluated.

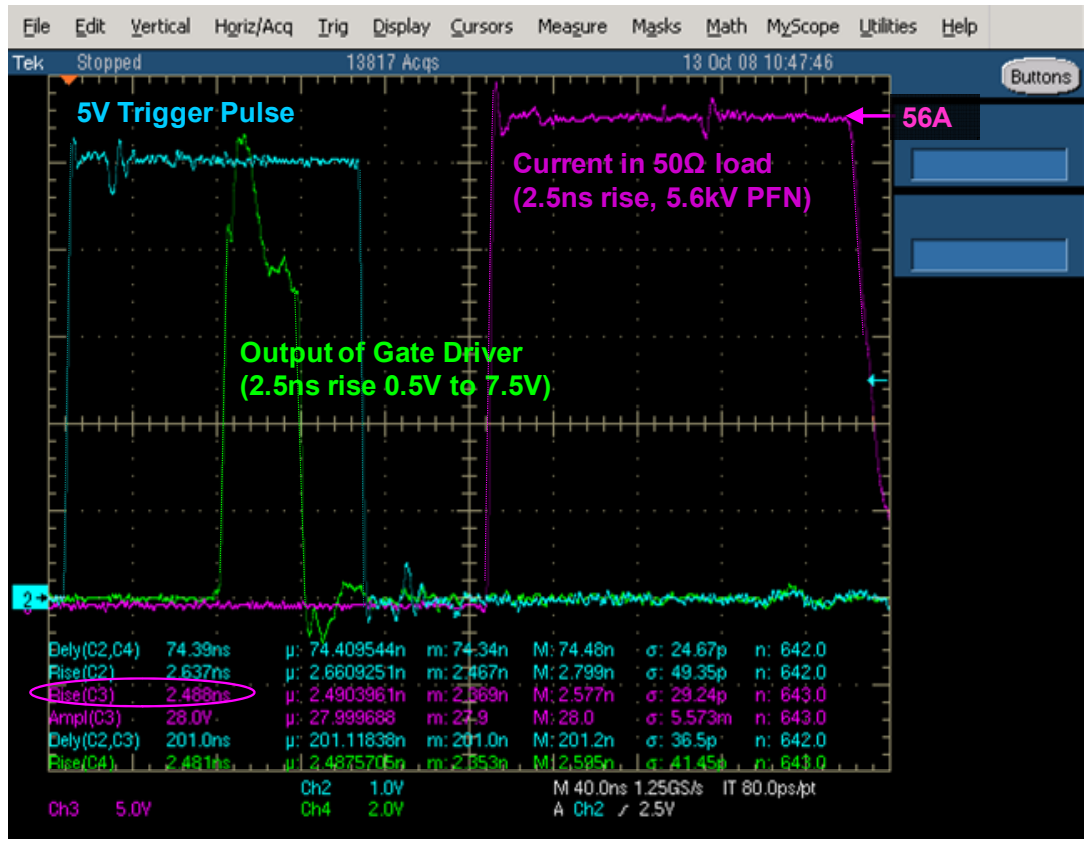

**Fig. 22:** Measured trigger pulse (cyan trace), gate driver output (green trace) and load current (lilac trace) for a CTF3 tail-clipper kicker system

**4.9 Resistive terminators**

In order that the impedance of the resistive terminator be matched to the system impedance, over a wide range of frequencies, high-power resistor disks are generally used, which are housed in a coaxial structure to minimize inductance of the terminator. The coaxial housing is normally tapered, with the maximum internal diameter of housing at the input end [49]: this design permits the resistive terminator to withstand a high pulse voltage while minimizing the parasitic inductance of the terminator.

Current distribution within the resistor discs, during the pulse, is dependent upon the resistivity of the disks and the frequency content of the pulse. As a result of proximity effect, image current flows on the inside diameter of the coaxial housing. For cooling and insulation purposes the terminator is generally filled with oil [18]. Long-term stability of the resistance value is linked to the ageing of the resistor discs: the ageing process is affected by the oil [33]. In order to accelerate the stabilization towards a final value of resistance, the discs are pre-impregnated under vacuum at high temperature. The short-term stability is linked to the temperature coefficient of the resistor material. Each current pulse will raise the temperature of the resistor stack: thus, where high stability is required, a suitable heat-exchanger is required to maintain the oil temperature.

The resistor discs also exhibit a voltage coefficient of resistance, which is typically around −1.5%/kV/cm: thus the resistive terminator changes in value during the pulse. For high-precision applications, where a well matched system is required, the voltage dependence of the terminator must be taken into account at the design stage.